\documentclass[superscriptaddress, aps,pre, twocolumn, floatfix]{revtex4-2}

\usepackage{graphicx}  
\usepackage{subfigure}
\usepackage[T1]{fontenc}
\usepackage{bm}        
\usepackage{amsfonts}  
\usepackage{amsmath}   
\usepackage{amssymb}   
\usepackage{bbold}
\usepackage{relsize}
\usepackage{url,hyperref}
\usepackage[table,usenames,dvipsnames]{xcolor}
\hypersetup{colorlinks=true, linkcolor=BrickRed,
  urlcolor=blue!50!black, citecolor=blue!50!black}

\newcommand{\pwisein}{\left\{ \begin{array}{ll}}
\newcommand{\pwiseout}{\end{array}\right.}

\newcounter{myequation}
\makeatletter
\@addtoreset{equation}{myequation}
\makeatother

\newcounter{myfigure}
\makeatletter
\@addtoreset{figure}{myfigure}
\makeatother

\begin{document}

\title{Cover time statistics of one-dimensional Brownian motion under stochastic resetting}
\author{Anirban Ghosh}
 \email{anirbansonapur@gmail.com}
 \affiliation{%
 Department of Physics, Amity Institute of Applied Sciences, Amity University Kolkata
}
 \affiliation{%
 Raman Research Institute, Bengaluru 560080, India
}
\author{Sanjib Sabhapandit}%
 \email{sanjib@rri.res.in}
\affiliation{%
 Raman Research Institute, Bengaluru 560080, India
}%

\date{\today}

\begin{abstract}

We investigate the effect of stochastic resetting on the statistics of the cover time of a one-dimensional Brownian motion with a diffusion constant $D$,  confined to a finite interval of length $L$. The cover time $t_c$, defined as the minimum time required for the particle to visit every point of the interval at least once, exhibits a non-monotonic dependence on the scaled reset rate $\rho = rL^2/4D$. The scaled mean cover time $4D\langle t_c\rangle/L^2$ initially decreases with increasing $\rho$,  attains a minimum at an optimal value $\rho^*$, and then increases with $\rho$, indicating an optimal resetting rate that minimizes the search duration. Furthermore, we derive an exact analytical expression for the cover time distribution, including its asymptotic limits, which agree well with numerical simulations. These results demonstrate that stochastic resetting serves as an efficient mechanism for optimizing cover-time processes in confined geometries.

\end{abstract}

\maketitle 

\section{Introduction}
Stochastic search processes are widespread in nature~\cite{RevModPhys8381}, encompassing activities like animals searching for food~\cite{edwards2007revisiting,viswanathan1996levy,viswanathan1999optimizing,viswanathan2011physics}, diverse biochemical reactions~\cite{shlesinger2009random,da2009random}, such as proteins seeking specific DNA sequences for binding~\cite{berg1981diffusion,mirny2008cell,gorman2008visualizing,gorman2010visualizing}, or sperm cells searching for an oocyte to fertilize\cite{eisenbach2006sperm,meerson2015mortality}. Many of these stochastic search processes are commonly represented by a single searcher executing a simple random walk (RW)\cite{RevModPhys8381,viswanathan1996levy,viswanathan2011physics,shlesinger2009random,da2009random,majumdar2016exact}. The primary metrics for assessing the effectiveness of such search processes have typically been expressed in terms of the time required for the searcher to reach a single target, commonly referred to as the first-passage time.
Several observables related to first-passage phenomena, such as the mean first-passage time\cite{redner2001guide, majumdar1999persistence}, survival probability at the origin\cite{majumdar1999persistence, redner2001guide,bray2013persistence,mejia2011first,bhat2016stochastic,ghosh2020persistence,ghosh2022persistence,mookerjee2025closed,chakraborty2007finite,chakraborty2008persistence,majumdar2017survival}, and the number of distinct and overlapping sites visited by random walkers\cite{larralde1992territory,acedo2003multiparticle,majumdar2012number,kundu2013exact,turban2014probability}, have been thoroughly investigated previously. 

However, when the task involves locating multiple targets, a situation frequently encountered in chemistry, ecology, or robotics~\cite{paramanick2024uncovering}, the relevant measure becomes the time required to reach a portion of the sites within the domain. The most comprehensive form of these exhaustive searches, where every site within a domain must be visited, defines what's known as the cover time. This metric is particularly significant as it signifies the time required to definitively locate all targets within a domain. The study of the cover time has been of considerable interest in random walk theory.

The search typically takes place within a limited domain where the targets are generally scattered across the entire area. As a result, identifying all these targets demands a comprehensive exploration of this bounded region. In this scenario, a key measure that indicates the efficiency of the search process is the cover time, $t_c$, which signifies the shortest time needed for the random walk (RW) to visit every location within the domain at least once~\cite{chupeau2015cover}. The cover time of a single random walker is also important in computer science, especially in the generation of random spanning trees (with uniform distribution) on any connected and undirected graph~\cite{aldous1990random}. Analytically determining the statistics of $t_c$ for a given confined domain remains a challenge in random walk (RW) theory. Prior research has largely focused on calculating the mean cover time on regular lattices, graphs, and networks~\cite{aldous1983time,yokoi1990some,broder1989bounds,brummelhuis1991covering,hemmer1998lattice,dembo2004cover,ding2012cover}. Obtaining analytical solutions for the entire distribution of $t_c$ is difficult and poses a significant challenge. In Ref.~\cite{zlatanov2009random}, the authors developed formal expressions for the distribution of $t_c$ on a general finite graph. Nevertheless, obtaining explicit results for large-scale systems from these expressions remains highly challenging. In Ref.~\cite{belius2013gumbel}, it was rigorously established that for random walks on a $d$-dimensional regular lattice with periodic boundary conditions (PBCs), the distribution of $t_c$ converges to a Gumbel distribution in the large-system limit, provided that $d > 2$ and the distribution is properly centered and scaled. Subsequently, it was demonstrated in Ref.~\cite{turban2015records} that the same conclusion applies to random walks on fully connected graphs, which formally corresponds to the limit $d\rightarrow\infty$. It's important to note that for $d$-dimensional regular graphs with $d>2$, the random walk (RW) is transient, meaning the walker escapes to infinity with a nonzero probability in the unbounded domain. Chupeau et al. showed that for any transient random walk on a finite graph, the appropriately centered and scaled cover time distribution robustly converges to a Gumbel distribution, independent of the graph topology~\cite{chupeau2015cover}.

A notable exception arises for one or two-dimensional random walks, where the walker is recurrent. In other words, when starting from a specific site, the walker inevitably returns to it with probability $1$. For $N$ walkers in one dimension, the cover time is defined as the minimum time required for all sites to be visited at least once by at least one walker. The cover time statistics of this problem were studied by Majumdar et al.~\cite{majumdar2016exact}.

In this paper, we investigate the effect of stochastic resetting on cover-time statistics for one-dimensional continuous Brownian motion confined by reflecting boundaries.
The theory of resetting processes has been developed considerably in the last decade~\cite{evans2011diffusion,kusmierz2014first,pal2015diffusion,campos2015phase,pal2016diffusion,ghosh2026anisotropic,reuveni2016optimal,evans2018run,nagar2023stochastic, evans2020stochastic, kundu2024preface,bressloff2020queueing,santra2022effect,meylahn2015large,majumdar2015dynamical,majumdar2022record,santra2020run,del2026proxitaxis,flaquer2024intermittent,nagar2023stochastic} and has allowed to gain understanding in a variety of topics such as the optimization of enzymatic reaction kinetics~\cite{reuveni2014role,rotbart2015michaelis}, computational searches~\cite{montanari2002optimizing} or animal foraging~\cite{boyer2014random,pal2020search,vilk2022phase}. The paradigmatic case of Brownian motion under stochastic resetting exhibits first-passage features different from ordinary diffusion~\cite{evans2011diffusion,reuveni2016optimal,pal2017first}. When the Brownian searcher is randomly reset to a fixed position with a constant reset rate, the mean first-passage time (MFPT) to a static target becomes finite
and can be minimized with respect to the reset rate~\cite{evans2011diffusion}. In fact, a key characteristic of a restart is
its potential to accelerate the completion of certain processes. For example, restarting can reduce the mean first-passage time of a diffusing particle to reach a boundary or target~\cite{reuveni2016optimal,evans2011diffusion,pal2017first,tal2020experimental,besga2020optimal,faisant2021optimal}, the mean run time of stochastic computer algorithms\cite{cayci2020continuous,lorenz2021restart,lorenz2018runtime}, the mean turnover time of
enzymatic reactions~\cite{reuveni2014role,rotbart2015michaelis}. Resetting typically intervenes by periodically halting the process at a specified rate and resetting it to an initial state, thereby limiting the process from pursuing inefficient, far-reaching trajectories. Although this induces unusual non-equilibrium behavior, it greatly boosts the efficiency of search processes by eliminating slower, extended paths~\cite{Pal2024,PhysRevE.99.032123,PhysRevE.103.052129,evans2011diffusion}. Resetting, however, is not necessarily beneficial in all cases, so a general criterion can be established to determine when it can effectively reduce the MFPT of a random walker to a static target~\cite{pal2017first, PhysRevLett.116.170601}. The effect of frequent resetting on the moments of cover time in $d$-dimensional spaces and on an arbitrary discrete network has been studied in the work by Linn et. al.~\cite{linn2025cover}.
We have investigated the effect of resetting on cover time $t_c$ for a continuous Brownian walker in one dimension. We have studied how the mean cover time and probability density function (PDF) change due to the application of resetting.

The paper is organized as follows. In Section~\ref{model_method}, we introduce the model of one-dimensional continuous Brownian motion confined between two reflecting boundaries and review the exact results for the cover time obtained by Majumdar et al.~\cite{majumdar2016exact}. Section~\ref{cover_reset_continuous} is devoted to the study of the system under stochastic resetting, where we modify the original model to incorporate resetting. We investigated the impact of the reset rate on the mean cover time through both analytical and numerical methods. Furthermore, we validated the full distribution of cover time under resetting including its asymptotic behavior through analytical derivations with simulation results. Finally, Section~\ref{conclusion} presents a brief summary of the results and discusses the relevance of the work.


\section{Model of Continuous Random walker in 1D with reflecting boundaries}\label{model_method}

Here, we recapitulate the framework and the main results of Majumdar et al.~\cite{majumdar2016exact} for the cover time of Brownian motion in a bounded domain of length $L$ in one dimension. 
We begin by considering a single Brownian particle confined within the finite interval $[0, L]$. The particle starts from an initial position $x_0$ and evolves under reflecting boundary conditions (RBC) at both $x = 0$ and $x = L$. The cover time, denoted by $t_c$, represents the time at which the particle has visited both boundaries for the first time. If $t_c > t$, it implies that at least one of the boundaries has not yet been reached by time $t$. Consequently, the probability that starting from the initial position $x_0$, the cover time exceeds $t$ can be expressed as, 
\begin{align}
\text{Prob.} [t_c>t|x_0] &= \text{Prob.} [L~\text{is unhit up to time}~t] \notag \\ &+ \text{Prob.} [0~\text{is unhit up to time}~t] \notag\\
&- \text{Prob.} [\text{both are unhit up to time}~t].
\end{align}

Each of these probabilities can be obtained by solving the standard backward Fokker–Planck equation with respect to the initial position $x_0$, for the survival probability $S(x_0, t)$ 
\begin{equation}
    \frac{\partial S(x_0, t)}{\partial t} = D\,\frac{\partial^2 S(x_0, t)}{\partial x_0^2}
    \label{e,q1}
\end{equation}
with the appropriate boundary conditions at $x_0=0$ and $x_0=L$. For instance, the probability that the point $L$ remains unhit up to time $t$ is denoted by $S_{AR}(x_0,t)$, where the subscript $A$ signifies an absorbing boundary condition at $x_0=L$ (i.e., $S(x_0=L,t)=0$), and the subscript $R$ signifies a reflecting boundary condition at $x_0=0$, (i.e., $\partial_{x_0}S(x_0,t)|_{x_0=0}=0$). Therefore, we have:
\begin{equation}
    \text{Prob.}[t_c>t|x_0]=S_{AR}(x_0,t)+S_{RA}(x_0,t)-S_{AA}(x_0,t)
    \label{eq2}
\end{equation}
where the subscripts refer to the boundary conditions. These survival probabilities have been computed exactly by Majumdar et. al.~\cite{majumdar2016exact}. In particular, for $x_0=L/2$, they found that
\begin{align}
    \label{eq3}
    S_{AA}(L/2,t)&=S_1(4Dt/L^2)\\
\label{eq4}
    S_{AR}(L/2,t)&=S_{AR}(L/2,t)=S_2(4Dt/L^2),
\end{align}
whose Laplace transforms $g_{n} (\lambda) = \int_0^\infty\, S_{n} (z)\, e^{-\lambda z}$ are given by 
\begin{equation}
    \begin{split}
        &g_1(\lambda)=\frac{1}{\lambda}\bigg[1-\frac{1}{\cosh\sqrt{\lambda}}\bigg]\\
        &g_2(\lambda)=\frac{1}{\lambda}\bigg[1-\frac{\cosh\sqrt{\lambda}}{\cosh2\sqrt{\lambda}}\bigg]
    \end{split}
    \label{eq:g-lambda}
\end{equation} 
These Laplace transforms can be inverted by using the standard Bromwich contour on the complex $\lambda$ plane and calculating the residues at the poles, yielding~\cite{majumdar2016exact},
\begin{equation}
    \begin{split}
       & S_1(z)=\frac{4}{\pi}\sum_{n=0}^{\infty}\frac{(-1)^n}{(2n+1)}e^{-(2n+1)^2\pi^2z/4}\\
       & S_2(z)=\frac{4}{\pi}\sum_{n=0}^{\infty}\frac{(-1)^n\cos{[(2n+1)\pi/4}]}{(2n+1)}e^{-(2n+1)^2\pi^2z/16}
    \end{split}
\end{equation}
The leading tail behaviors of $S_1(z)$ and $S_2(z)$ are given by the $n=0$ terms in the corresponding series as,~\cite{majumdar2016exact}
\begin{equation}
\begin{split}
&S_1(z)\underset{z \to \infty}{\approx}  \frac{4}{\pi}e^{-\pi^2z/4}\\
&S_2(z)\underset{z \to \infty}{\approx} \frac{4}{\pi\sqrt{2}}e^{-\pi^2z/16}
\end{split}
\end{equation}

For $x_0=L/2$, let us denote the 
cumulative probability distribution and the probability density function of the cover time as $R(t)$ and $p(t_c)$ respectively. Therefore
\begin{equation}
    R(t) = \int_{t}^{\infty} p(t_c)\, dt_c= S_{AR}(t) + S_{RA}(t) - S_{AA}(t),
    \label{eq:CDFx10}
\end{equation}
where we have omitted the $L/2$ in the arguments.  
The mean cover time can then be written as
\begin{equation}
    \langle t_c \rangle 
    = \int_{0}^{\infty} t\, p(t)\, dt 
    = \int_{0}^{\infty}  R(t)\, dt 
    = \widetilde{R}(s=0),
\label{cov}
\end{equation}
where \( \widetilde{R}(s) \) is the Laplace transform of \( R(t) \), and we have used integration by parts. Taking a Laplace transform of Eq.~\eqref{eq:CDFx10}, the mean cover time is then given by
\begin{equation}
\langle t_c \rangle = 
    \widetilde{R}(0)
    = \widetilde{S}_{AR}(0)
    + \widetilde{S}_{RA}(0)
    - \widetilde{S}_{AA}(0).
\label{14}
\end{equation}


\section{Continuous Random walker in 1D with
reflecting boundaries under resetting}\label{cover_reset_continuous}

\begin{figure}
    \centering
    \includegraphics[width=0.5\textwidth]{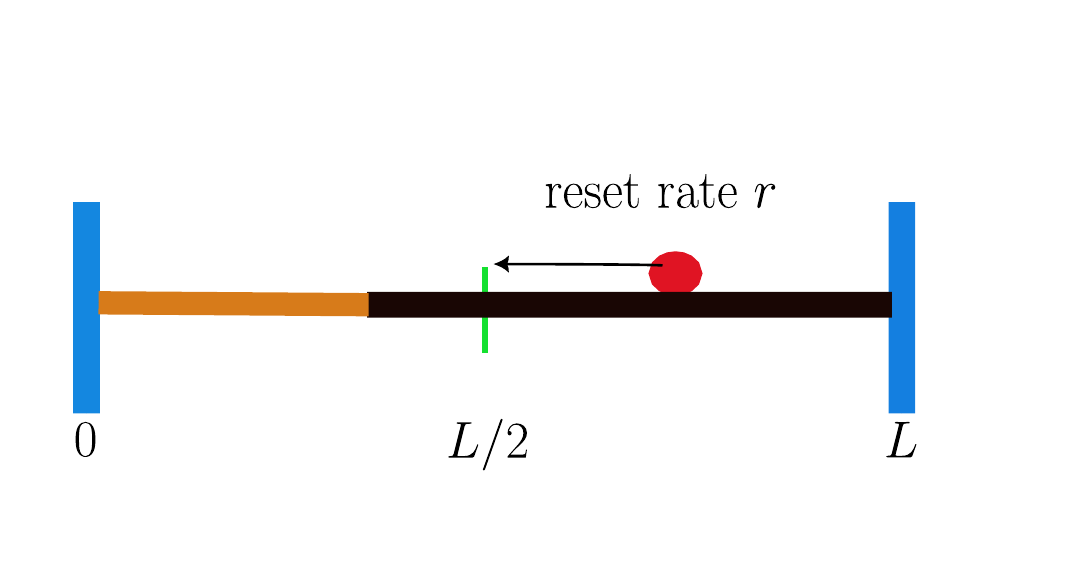}
    \caption{Brownian motion (BM) with reflecting boundary condition at $x=0$ and $x=L$, which are shown in blue. The black region indicates the space already covered by the walker up to time $t$, starting at $x_0=L/2$. The Orange region is yet to be covered. The Brownian particle is resetting at $L/2$ with a rate $r$. The cover time $t_c$ is the first time at which the entire domain becomes black.}
    \label{fig_reset}
\end{figure}

\subsection{Renewal equation approach for Poissonian resetting}
First, we define diffusion with Poissonian resetting in one dimension. We consider a single Brownian particle moving on the real line, starting from the initial position $x_0 = L/2$ at time $t=0$. The particle undergoes stochastic resetting to its initial position $X_r = L/2$ with a constant rate $r$, as illustrated schematically in Fig.~(\ref{fig_reset}). 

The dynamics thus consist of alternating phases of free diffusion and instantaneous resetting to the fixed location $L/2$. The position $x(t)$ of the particle at time $t$ evolves according to the following stochastic rule \cite{evans2011diffusion}. During a small time interval $dt$, the position $x(t)$ is updated to
\[
x(t + dt) =
\begin{cases}
L/2 & \text{with probability } r \, dt, \\
x(t) +  \sqrt{2Ddt}\, \eta(t)  & \text{with probability } 1 - r \, dt.
\end{cases}
\]
where $\{\eta(t)\}$ (in steps of $dt$) are independent and identically distributed Gaussian random variables with zero mean and unit variance.  
For Poissonian resetting, for a generic process, it is possible to relate the survival probability with resetting, $Q_r$, to that without resetting, $Q_0$, by using a last renewal equation, which reads~\cite{majumdar2015dynamical}
\begin{align}
Q_r(x_0, t) &= e^{-rt} Q_0(x_0, t) \notag \\&+ r \int_0^t d\tau \, e^{-r\tau} Q_0(X_r, \tau) Q_r(x_0, t - \tau)
\label{renewal}
\end{align}
The first term in Eq.~(\ref{renewal}) represents trajectories in which there has been no resetting. The second term represents trajectories in which at least one resetting event has occurred.  The integral is over $\tau$,  the time elapsed since the last reset, and we have a convolution of survival probabilities: survival starting from $x_0$ with resetting up to time $t-\tau$ (the time of the last reset) and survival starting from $X_r$ in the absence of resetting for duration $\tau$. In the Laplace space, 
\begin{math}
    \Tilde{Q}_r(x_0,s)=\int_{0}^{\infty} dt e^{-st}Q_r(x_0,t),
\end{math}
Eq.(\ref{renewal}) leads to the expression~\cite{majumdar2015dynamical}
\begin{equation}
    \Tilde{Q}_r(x_0,s)=\frac{\Tilde{Q}_0(x_0,r+s)}{1-r\Tilde{Q}_0(X_r,r+s)}
    \label{res}
\end{equation}
This is a very general result for Poissonian resetting, which relates the Laplace transform of the survival probability with resetting to that without resetting. This result will be useful for the following.


\subsection{Cover time under resetting and optimal reset rate} 

We examine how resetting influences cover time statistics and its efficacy in streamlining search durations. Given that the cover time is comprised of two distinct survival probability components, $\tilde{S}_{AR}(s)$ and $\tilde{S}_{AA}(s)$, as shown in Eq.~(\ref{14}), we analyze the impact of resetting on each independently. By incorporating the standard resetting framework into Eq.~(\ref{res}) for both $\tilde{S}_{AR}(s)$ and $\tilde{S}_{AA}(s)$, we observe that
\begin{equation}
\begin{split}
&\Tilde{S}_{AA}^{r}(s)=\frac{\Tilde{S}_{AA}(r+s)}{1-r\Tilde{S}_{AA}(r+s)},\\
    &\Tilde{S}_{AR}^{r}(s)=\frac{\Tilde{S}_{AR}(r+s)}{1-r\Tilde{S}_{AR}(r+s)}.
    \end{split}
    \label{SAAAR}
\end{equation}
It follows from Eqs.~\eqref{eq3}--\eqref{eq4} that 
\begin{equation}
    \tilde{S}_{AA} (s)= \frac{L^2}{4 D} g_1\left(\frac{L^2s}{4D}\right)~~\text{and}~~\tilde{S}_{AR} (s)= \frac{L^2}{4 D} g_2\left(\frac{L^2s}{4D}\right).
    \label{eq:S-scal}
\end{equation}
Using the expressions of $g_1(\lambda)$ and $g_2(\lambda)$ from Eq.~\eqref{eq:g-lambda}, from Eqs.~\eqref{SAAAR}--\eqref{eq:S-scal}, we now get \begin{equation}
        \Tilde{S}_{AA}^{r}(s)
        =\frac{\cosh{\sqrt{\frac{L^2(r+s)}{4D}}}-1}{s\cosh{\sqrt{\frac{L^2(r+s)}{4D}}}+r},
    \label{saa}
\end{equation}
and,
\begin{equation}
    \Tilde{S}_{AR}^{r}(s)=\frac{\cosh{2\sqrt{\frac{L^2(r+s)}{4D}}}-\cosh{\sqrt{\frac{L^2(r+s)}{4D}}}}{s\cosh{2\sqrt{\frac{L^2(r+s)}{4D}}}+r\cosh{\sqrt{\frac{L^2(r+s)}{4D}}}}.
    \label{sar}
\end{equation}
To calculate the mean cover time, we set $s=0$, which yields
\begin{equation}
        \Tilde{S}_{AA}^{r}(0)
        =\frac{1}{r}\Bigg[\cosh\sqrt{\frac{rL^2}{4D}}-1\Bigg],
    \label{27}
\end{equation}
and
\begin{equation}
        \Tilde{S}_{AR}^{r}(0)=\frac{1}{r}\Bigg[\frac{\cosh2\sqrt{\frac{rL^2}{4D}}}{\cosh\sqrt{\frac{rL^2}{4D}}}-1\Bigg].
\label{28}
\end{equation}

Thus, the mean cover time $\langle t_c\rangle$ under resetting can be exactly calculated by substituting the expressions from Eqs.~(\ref{27}) and (\ref{28}) into Eq.~(\ref{14}), which, after simplification,  yields
\begin{equation}
        \langle t_c\rangle_r
        =\frac{L^2}{4D} f\left(\frac{rL^2}{4D}\right),
    \label{mean_cover_time}
\end{equation}
where the scaling function $f(\rho)$ is given by
\begin{equation}
    f(\rho)=\frac{1}{\rho}\Big(3\cosh{\sqrt{\rho}}-\frac{2}{\cosh{\sqrt{\rho}}}-1\Big).
    \label{diff}
\end{equation}

\begin{figure}
    \centering
\includegraphics[width=0.5\textwidth]{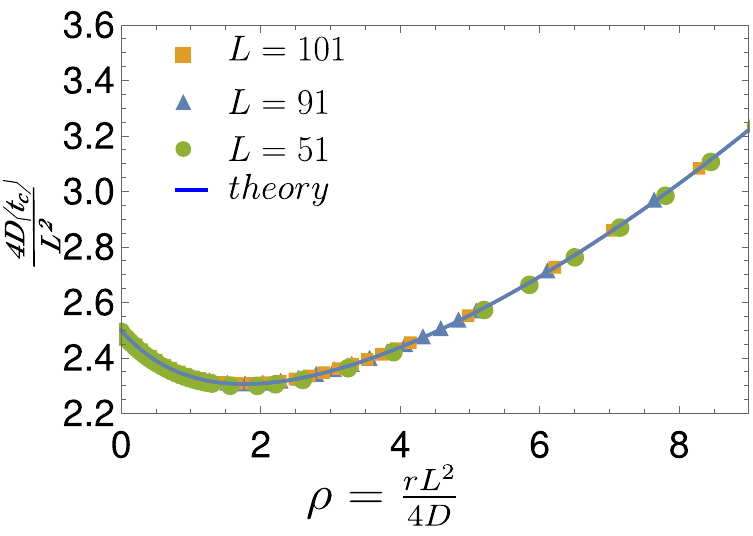}
    \caption{Simulation results of scaled mean cover time with scaled resetting rate $\rho=rL^2/4D$ for different values of $L$ shown in the figure. $D$ is kept fixed at $0.5$. The solid line represents the analytical result found in Eq.(\ref{mean_cover_time}) for different values of $L$. The minima is calculated at the scaled reset rate $\rho^*=1.76976$, and the minimum scaled mean cover time is found as $2.30625$.}
    \label{fig1}
\end{figure}

In the limit $\rho\to 0$, from Eq.~\eqref{diff}, one finds 
\begin{equation}
 f(\rho)= \frac{5}{2}-\frac{7 \rho}{24} + O (\rho^2). 
\end{equation}
On the other hand, $f(\rho)$ diverges exponentially, $f(\rho)\sim e^{\sqrt{\rho}}$, as $\rho\to \infty$. Since the slope $f'(0) = -7/24 <0$, it indicates that the function $f(\rho)$ has at least one minimum. Indeed, as shown in Fig.~\ref{fig1}, the scaled mean cover time $f(\rho)$ has a single minimum as a function of the scaled resetting rate $\rho=r L^2/(4D)$.

To find the location of the minimum,
we differentiate the function in Eq.~(\ref{diff}) with respect to $\rho$ and set the derivative to zero. The derivative of $f(\rho)$ from Eq.~(\ref{diff}) is given by
\begin{equation}
\begin{split}
f'(\rho) &= \rho^{-3/2} \left( \frac{3\sinh{\sqrt{\rho}}}{2} + \frac{\sinh{\sqrt{\rho}}}{\cosh^2{\sqrt{\rho}}} \right) \\
&\quad - \frac{1}{\rho^2} \left( 3\cosh{\sqrt{\rho}} - \frac{2}{\cosh{\sqrt{\rho}}} - 1 \right).
\end{split}
\label{opti}
\end{equation}
Solving $f'(\rho) = 0$ gives the location of the minimum at 
\begin{equation}
\rho^* = 1.76976\dots.
\label{eq:rhostar}
\end{equation}
Consequently, the optimal resetting rate is given by 
\begin{equation}
    r^* = \tau_d^{-1} \times 7.07904\dots\,,
\end{equation}
where $\tau_d = L^2/D$ denotes the diffusive time scale. 

Substituting $\rho^*$ from Eq.~\eqref{eq:rhostar} in Eq.~(\ref{mean_cover_time}) yields the minimum scaled mean cover time:
\begin{equation}
   \frac{4D\langle t_c\rangle_{\min}}{L^2} = 2.30625\dots. 
\end{equation}

We now discuss the simulation results for the mean cover time of a one-dimensional continuous Brownian motion with resetting. We investigate the effect of resetting in Fig.~\ref{fig1} for different system sizes $L$. We used the discrete version of the Langevin equation to update the position for each time step $dt=0.001$. In our simulation, we continue the Brownian motion until both boundaries are hit, and note the cover time for each configuration. We use $10^7$ realizations to find the mean cover time. In Fig.~\ref{fig1}, we plot the scaled mean cover time $4D\langle t_c\rangle/L^2$ with the scaled reset rate $\rho=rL^2/(4D)$. We find that the scaled mean cover time decreases with increasing scaled reset rate and attains a minimum at $\rho^*$. As seen in Fig.~\ref{fig1}, the theoretical prediction of Eq.~(\ref{diff}) is in excellent agreement with numerical simulation.


\subsection{ Distribution of cover time under resetting}

In the absence of any resetting, the probability density function (PDF) of the cover time is given by~\cite{majumdar2016exact}, 
\begin{equation}
    p(t_c) = \frac{4D}{L^2}\, f_1^{0}\left(\frac{4 D t_c}{L^2}\right),
\end{equation}
with the scaling function $f_1^0(z)$ given by~\cite{majumdar2016exact},
\begin{align}
         f_1^{0}(z)=& \frac{\pi}{2}\sum_{n=0}^{\infty} (-1)^n(2n+1)\notag\\ &\times \Big[\cos{[(2n+1)\pi/4]}\, e^{-(2n+1)^2\pi^2z/16}\notag \\
        &\qquad -2\, e^{-(2n+1)^2\pi^2z/4}\Big].
    \label{pdf1}
\end{align}

Now we want to find out how the PDF in Eq.~\eqref{pdf1} is modified in the presence of a nonzero resetting rate $r$. Following Eq.~\eqref{eq:CDFx10},  the cumulative distribution of the cover time, in the presence of resetting, is given by 
\begin{equation}
    \text{Prob.}\, [t_c>t|L]  = 2 S^r_{AR}(t) - S^r_{AA}(t),
\end{equation}
where the Laplace transforms of $S^r_{AA}(t)$ and $S^r_{AR}(t)$ are given in  Eqs.(\ref{saa}) and (\ref{sar}) respectively. From the expressions of the Laplace transforms, it is clear that 
\begin{align}
    \label{eq3x}
    S^r_{AA}(L/2,t)&=S_1^r(4Dt/L^2)\\
\label{eq4x}
    S^r_{AR}(L/2,t)&=S_{AR}^r(L/2,t)=S_2(4Dt/L^2),
\end{align}
where $S_1^r(z)$ and $S_2^r(z)$ depends on the resetting rate $r$ only through the scaled parameter $\rho=rL^2/(4D)$. Therefore, 
\begin{equation}
    \text{Prob.}\, [t_c>t|L] = F_1^r\left(\frac{4D t}{L^2}\right),
\end{equation}
with
\begin{equation}
    F_1^r(z) = 2 S_2^r(z) - S_1^r (z),
    \label{eq:F1r}
\end{equation}

\begin{figure}[t!]
    \centering
    \includegraphics[width=0.5\textwidth]{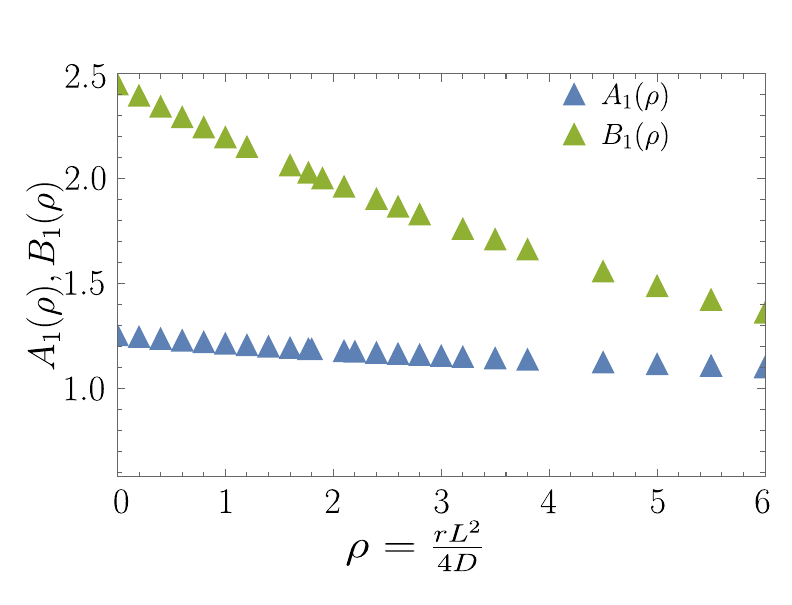}\\
    \includegraphics[width=0.5\textwidth]{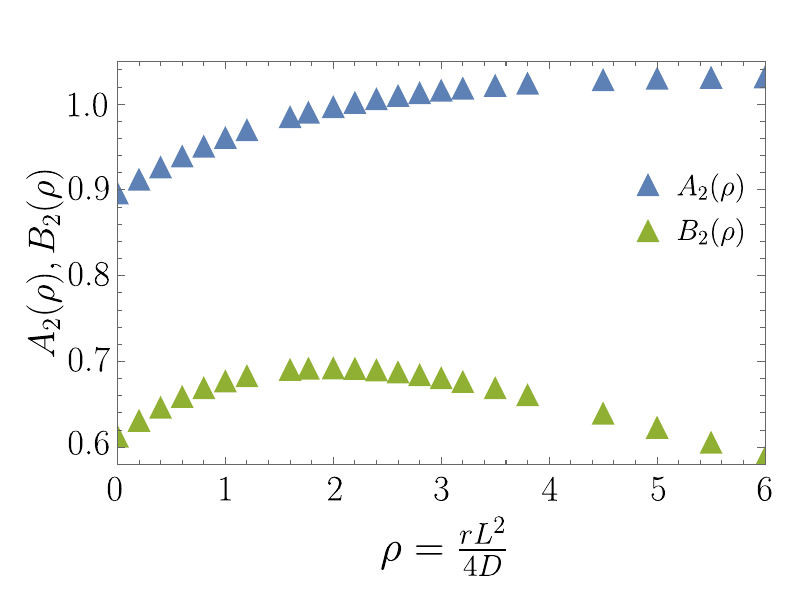}
    \caption{Plot of $A_{1,2}(\rho)$ and $B_{1,2}(\rho)$, obtained by using the expressions \eqref{eq:Brho}--\eqref{eq:Arho}. }
    \label{fig5}
\end{figure}

We find that [see Appendix~(\ref{app}) for details],  the leading tail behavior of $S_{1,2}^r(z)$ for large $z$, are given by  
\begin{equation}
    \begin{split}
        & S_1^r(z)\sim A_1(\rho)e^{-B_1(\rho)z}\\
        & S_2^r(z)\sim A_2(\rho)e^{-B_2(\rho)z}
    \end{split}
    \label{eq:S12rx}
\end{equation}
where 
\begin{equation}
  B_1(\rho) = [\rho - w_1(\rho)]\quad \text{and} \quad B_2(\rho) = [\rho - w_2(\rho)],
  \label{eq:Brho}
\end{equation} 
and  
\begin{widetext}
\begin{equation}
\begin{split}&A_1(\rho)=\frac{2\sqrt{w_1(\rho)}\Big(\cosh{(\sqrt{w_1(\rho)})}-1\Big)}{2\sqrt{w_1(\rho)}\cosh{(\sqrt{w_1(\rho)})}+(w_1(\rho)-\rho)\sinh{(\sqrt{w_1(\rho)}})}\\
    &A_2(\rho)=\frac{2\sqrt{w_2(\rho)}\Big(\cosh{(2\sqrt{w_2(\rho)})}-\cosh{(\sqrt{w_2(\rho)})}\Big)}{\rho\sinh{(\sqrt{w_2(\rho)})}+2(w_2(\rho)-\rho)\sinh{(2\sqrt{w_2(\rho)})}+2\sqrt{w_2(\rho)}\cosh{(2\sqrt{w_2(\rho)})}}
    \end{split}
    \label{eq:Arho}
\end{equation}
\end{widetext}
The functions $w_{1,2}(\rho)$ are the negative roots closest to the $w_{1,2}=0$, 
of the equations 
\begin{equation}
   (w_1-\rho)\cosh{\sqrt{w_1}}+\rho =0  
\end{equation}
and
\begin{equation}
    (w_2-\rho)\cosh{2\sqrt{w_2}}+\rho \cosh{\sqrt{w_2}} =0
\end{equation}
respectively, for a given value of $\rho$. The functions $A_{1,2}(\rho)$ and $B_{1,2}(\rho)$ are shown in Fig.~\ref{fig5}.

\begin{figure}
    \centering
    \includegraphics[width=0.5\textwidth]{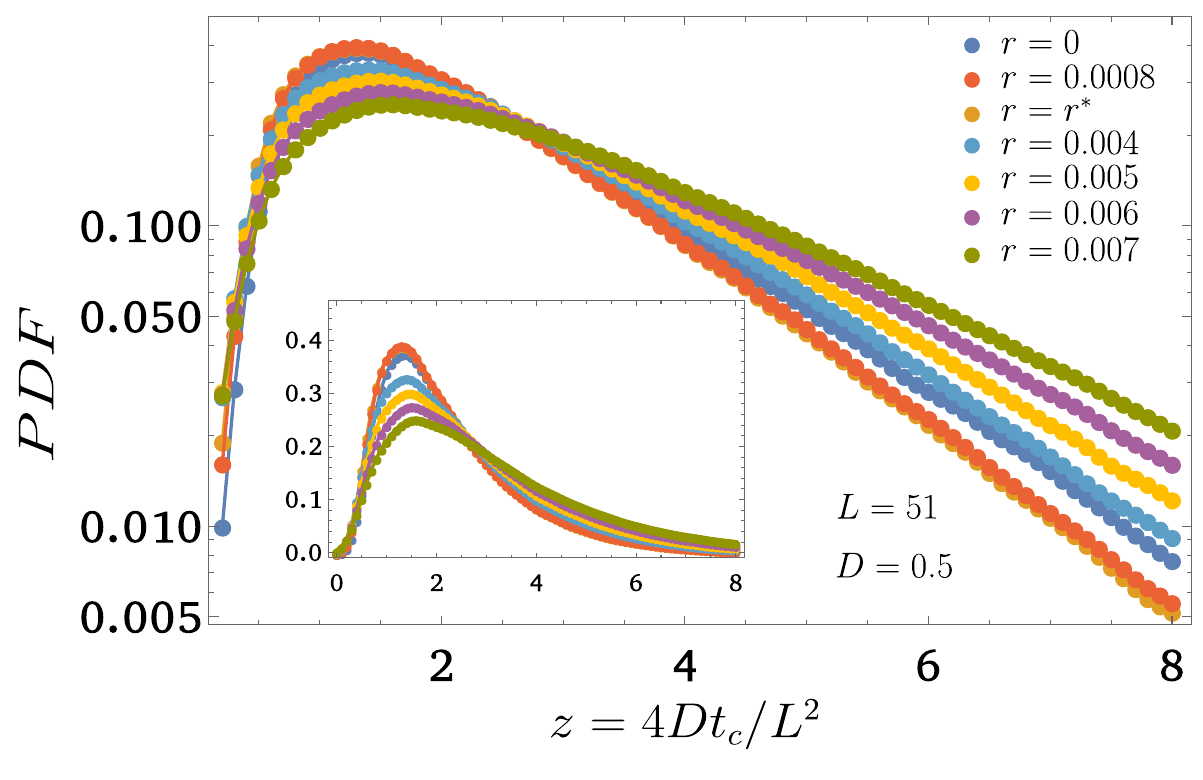}
    \caption{The PDFs of the scaled cover time for a continuous Brownian motion with RBC in log-linear scale for different values of resetting rate $r$ at $D=0.5$. The solid lines are representations of theoretical expression of the distribution function written in Eqs.(\ref{exact_pdf}) and (\ref{dist_small_z}) and closed circles are the representation of the simulation results for different values of $r$ written in the legends for $L=51$. The inset: the PDF is shown in linear scale.}
    \label{fig2}
\end{figure}

Finally, using using $p^r(t_c) = - \frac{d}{dt_c} F_1^r (4 D t_c/L^2)$, gives the PDF of the cover time with resetting as
\begin{equation}
    p^r(t_c) = \frac{4D}{L^2}\, f_1^{r}\left(\frac{4 D t_c}{L^2}\right),
\end{equation}
where from Eq.~\eqref{eq:F1r}, the scaling function $f_1^r(z)$ is given by
\begin{equation}
    f_1^r(z)= \frac{d}{dz} \left[S_1^r(z) -2 S_2^r(z)\right]. 
\end{equation}
Using the explicit forms of $S_{1,2}^r(z)$ from Eq.~\eqref{eq:S12rx} yields, 
\begin{equation}
    \begin{split}
        f_1^r(z)\sim 2A_2(\rho)B_2(\rho)e^{-B_2(\rho)z}-A_1(\rho)B_1(\rho)e^{-B_1(\rho)z},
    \end{split}
    \label{exact_pdf}
\end{equation}
for large $z$. On the other hand, the small $z$ leading behavior of $f_1^r(z)$ can be found as [see Appendix~(\ref{appB}) for details],
\begin{equation}
    f_1^r(z,\rho) = \frac{6}{\sqrt{\pi}} \, z^{-3/2} \, e^{-\left(z \rho + \frac{9}{4z}\right)} 
    \label{dist_small_z}
\end{equation}
Fig.~\ref{fig2} shows an excellent agreement between the analytical prediction and numerical simulations.

\section{Conclusion} \label{conclusion}

In summary, we have investigated the cover time statistics of a Brownian particle on a one-dimensional line of length $[0,L]$ with reflective boundaries in the presence of resetting dynamics. We have explored the multiple-target search problem by analyzing a Brownian particle influenced by thermal noise within the framework of an instantaneous stochastic resetting mechanism. The scaled mean cover time  $\frac{4D\langle t_c \rangle}{L^2}$ was found to reach its minimum value of \( 2.30625 \) at a scaled reset rate of \( \rho^* = 1.76976 \). Additionally, we explored how the scaled cover time distribution function varies with different reset rates. The functional form of the scaling function was derived, encompassing expressions for both small and large values of the scaled cover time. Since resetting is known to optimize search time in stochastic processes, and cover time is an important observable for exhaustive searches, our results on cover time statistics with resetting provide a valuable starting point and foundation for future research in this area.


\bibliography{reference}

@article{RevModPhys8381,
  title={Intermittent search strategies},
  author = {B\'enichou, O. and Loverdo, C. and Moreau, M. and Voituriez, R.},
  journal = {Rev. Mod. Phys.},
  volume = {83},
  issue = {1},
  pages = {81--129},
  numpages = {0},
  year = {2011},
  month = {Mar},
  publisher = {American Physical Society},
  doi = {10.1103/RevModPhys.83.81},
  url = {https://link.aps.org/doi/10.1103/RevModPhys.83.81}
}

@article{edwards2007revisiting,
  title={Revisiting L{\'e}vy flight search patterns of wandering albatrosses, bumblebees and deer},
  author={Edwards, Andrew M and Phillips, Richard A and Watkins, Nicholas W and Freeman, Mervyn P and Murphy, Eugene J and Afanasyev, Vsevolod and Buldyrev, Sergey V and Stanley, H Eugene and others},
  journal={Nature},
  volume={449},
  number={7165},
  pages={1044--1048},
  year={2007},
  publisher={Nature Publishing Group UK London},
  url = {https://www.nature.com/articles/nature06199}
}

@article{viswanathan1996levy,
title={L{\'e}vy flight search patterns of wandering albatrosses},
  author={Viswanathan, Gandhimohan M and Afanasyev, Vsevolod and Buldyrev, Sergey V and Murphy, Eugene J and Prince, Peter A and Stanley, H Eugene},
  journal={Nature},
  volume={381},
  number={6581},
  pages={413--415},
  year={1996},
  publisher={Nature Publishing Group UK London},
  url = {https://www.nature.com/articles/381413a0}
}

@article{viswanathan1999optimizing,
title={Optimizing the success of random searches},
  author={Viswanathan, Gandimohan M and Buldyrev, Sergey V and Havlin, Shlomo and da Luz, Marcos GE and Raposo, Ernesto P and Stanley, H Eugene},
  journal={nature},
  volume={401},
  number={6756},
  pages={911--914},
  year={1999},
  publisher={Nature Publishing Group UK London},
  url={https://www.nature.com/articles/44831}
}

@book{viswanathan2011physics,
title={The Physics of Foraging: An Introduction to Random Searches and Biological Encounters},
  author={Viswanathan, Gandhimohan M and Da Luz, Marcos GE and Raposo, Ernesto P and Stanley, H Eugene},
  year={2011},
  publisher={Cambridge University Press},
  url = {https://www.cambridge.org/core/books/physics-of-foraging B009DE42189D3A39718C2E37EBE256B0}
}

@article{shlesinger2009random,
title={Random searching},
  author={Shlesinger, Michael F},
  journal={J. Phys. A: Math. Theor.},
  volume={42},
  number={43},
  pages={434001},
  year={2009},
  publisher={IOP Publishing},
  url={https://iopscience.iop.org/article/10.1088/1751-8113/42/43/434001}
}

@article{da2009random,
title={The random search problem: trends and perspectives},
  author={Da Luz, Marcos GE and Grosberg, Alexander and Raposo, Ernesto P and Viswanathan, Gandhi M},
  journal={J. Phys. A: Math. Theor.},
  volume={42},
  number={43},
  pages={430301},
  year={2009},
  publisher={IOP Publishing},
  url={https://iopscience.iop.org/article/10.1088/1751-8121/42/43/430301}
}

@article{berg1981diffusion,
title={Diffusion-driven mechanisms of protein translocation on nucleic acids. 1. Models and theory},
  author={Berg, Otto G and Winter, Robert B and Von Hippel, Peter H},
  journal={Biochemistry},
  volume={20},
  number={24},
  pages={6929--6948},
  year={1981},
  publisher={ACS Publications},
  url={https://pubs.acs.org/doi/10.1021/bi00527a028}
}

@article{mirny2008cell,
title={Cell commuters avoid delays},
  author={Mirny, Leonid},
  journal={Nat. Phys.},
  volume={4},
  number={2},
  pages={93--95},
  year={2008},
  publisher={Nature Publishing Group UK London},
  url={https://www.nature.com/articles/nphys848}
}

@article{gorman2008visualizing,
  title={Visualizing one-dimensional diffusion of proteins along DNA},
  author={Gorman, Jason and Greene, Eric C},
  journal={Nat. Struct. Mol. Biol.},
  volume={15},
  number={8},
  pages={768--774},
  year={2008},
  publisher={Nature Publishing Group US New York},
  url={https://www.nature.com/articles/nsmb.1441}
}

@article{gorman2010visualizing,
  title={Visualizing one-dimensional diffusion of eukaryotic DNA repair factors along a chromatin lattice},
  author={Gorman, Jason and Plys, Aaron J and Visnapuu, Mari-Liis and Alani, Eric and Greene, Eric C},
  journal={Nat. Struct. Mol. Biol.},
  volume={17},
  number={8},
  pages={932--938},
  year={2010},
  publisher={Nature Publishing Group US New York},
  url={https://www.nature.com/articles/nsmb.1858}
}

@article{eisenbach2006sperm,
  title={Sperm guidance in mammals — an unpaved road to the egg},
  author={Eisenbach, Michael and Giojalas, Laura C},
  journal={Nat. Rev. Mol. Cell Biol.},
  volume={7},
  number={4},
  pages={276--285},
  year={2006},
  publisher={Nature Publishing Group UK London},
  url={https://www.nature.com/articles/nrm1893}
}

@article{meerson2015mortality,
  title = {Mortality, Redundancy, and Diversity in Stochastic Search},
  author = {Meerson, Baruch and Redner, S.},
  journal = {Phys. Rev. Lett.},
  volume = {114},
  issue = {19},
  pages = {198101},
  numpages = {5},
  year = {2015},
  month = {May},
  publisher = {American Physical Society},
  doi = {10.1103/PhysRevLett.114.198101},
  url = {https://link.aps.org/doi/10.1103/PhysRevLett.114.198101}
}

@article{chupeau2015cover,
  title={Cover times of random searches},
  author={Chupeau, Marie and B{\'e}nichou, Olivier and Voituriez, Rapha{\"e}l},
  journal={Nat. Phys.},
  volume={11},
  number={10},
  pages={844--847},
  year={2015},
  publisher={Nature Publishing Group UK London},
  url={https://www.nature.com/articles/nphys3413}
}

@article{majumdar2016exact,
  title={Exact distributions of cover times for $N$ independent random walkers in one dimension},
  author={Majumdar, Satya N and Sabhapandit, Sanjib and Schehr, Gr{\'e}gory},
  journal={Phys. Rev. E},
  volume={94},
  number={6},
  pages={062131},
  year={2016},
  publisher={APS},
  url={https://journals.aps.org/pre/abstract/10.1103/PhysRevE.94.062131}
}

@article{aldous1990random,
title={The Random Walk Construction of Uniform Spanning Trees and Uniform Labelled Trees},
  author={Aldous, David J},
  journal={SIAM J. Discrete Math.},
  volume={3},
  number={4},
  pages={450--465},
  year={1990},
  publisher={SIAM},
  url={https://epubs.siam.org/doi/10.1137/0403039}
}

@article{aldous1983time,
title={On the time taken by random walks on finite groups to visit every state},
  author={Aldous, David J},
  journal={Zeitschrift f{\"u}r Wahrscheinlichkeitstheorie und verwandte Gebiete},
  volume={62},
  number={3},
  pages={361--374},
  year={1983},
  publisher={Springer},
  url={https://link.springer.com/article/10.1007/BF00535260}
}

@article{broder1989bounds,
  title={Bounds on the cover time},
  author={Broder, Andrei Z and Karlin, Anna R},
  journal={J. Theor. Probab.},
  volume={2},
  pages={101--120},
  year={1989},
  publisher={Springer},
  url={https://link.springer.com/article/10.1007/BF01048273}
}

@article{yokoi1990some,
  title={Some exact results for the lattice covering time problem},
  author={Yokoi, Carlos SO and Hern{\'a}ndez-Machado, A and Ram{\'\i}rez-Piscina, L},
  journal={Phys. Lett. A},
  volume={145},
  number={2-3},
  pages={82--86},
  year={1990},
  publisher={Elsevier},
  url={https://doi.org/10.1016/0375-9601(90)90196-U}
}

@article{brummelhuis1991covering,
  title={Covering of a finite lattice by a random walk},
  author={Brummelhuis, MJAM and Hilhorst, HJ},
  journal={Physica A},
  volume={176},
  number={3},
  pages={387--408},
  year={1991},
  publisher={Elsevier},
  url={https://doi.org/10.1016/0378-4371(91)90220-7}
}

@article{hemmer1998lattice,
title={Lattice covering by two random walkers in one dimension},
  author={Hemmer, PC and Hemmer, S},
  journal={Physica A},
  volume={251},
  number={1-2},
  pages={245--250},
  year={1998},
  publisher={Elsevier},
  url={https://doi.org/10.1016/S0378-4371(97)00608-0}
}

@article{dembo2004cover,
title={Cover Times for Brownian Motion and Random Walks in Two Dimensions},
  author={Dembo, Amir and Peres, Yuval and Rosen, Jay and Zeitouni, Ofer},
  journal={Ann. Math.},
  pages={433--464},
  year={2004},
  publisher={JSTOR},
  url={https://www.jstor.org/stable/3597219}
}

@article{ding2012cover,
title={On cover times for 2D lattices},
  author={Ding, Jian},
  journal={Electron. J. Probab.},
  volume={17},
  number={45},
  pages={1--18},
  year={2012},
  publisher={Citeseer},
  url={https://projecteuclid.org/journals/electronic-journal-of-probability/volume-17/issue-none/On-cover-times-for-2D-lattices/10.1214/EJP.v17-2089.full}
}

@article{zlatanov2009random,
title={Random walks on networks: Cumulative distribution of cover time},
  author={Zlatanov, Nikola and Kocarev, Ljupco},
  journal={Phys. Rev. E},
  volume={80},
  number={4},
  pages={041102},
  year={2009},
  publisher={APS},
  url={https://doi.org/10.1103/PhysRevE.80.041102}
}

@article{belius2013gumbel,
title={Gumbel fluctuations for cover times in the discrete torus},
  author={Belius, David},
  journal={Probab. Theory Relat. Fields},
  volume={157},
  pages={635--689},
  year={2013},
  publisher={Springer},
  url={https://link.springer.com/article/10.1007/s00440-012-0467-7}
}

@article{turban2015records,
title={Records for the number of distinct sites visited by a random walk on the fully connected lattice},
  author={Turban, Lo{\"\i}c},
  journal={J. Phys. A: Math. Theor.},
  volume={48},
  number={44},
  pages={445001},
  year={2015},
  publisher={IOP Publishing},
  url={https://iopscience.iop.org/article/10.1088/1751-8113/48/44/445001}
}

@book{redner2001guide,
  title={A guide to first-passage processes},
  author={Redner, Sidney},
  year={2001},
  publisher={Cambridge university press}
}

@article{bray2013persistence,
title={Persistence and first-passage properties in nonequilibrium systems},
  author={Bray, Alan J and Majumdar, Satya N and Schehr, Gr{\'e}gory},
  journal={Adv. Phys.},
  volume={62},
  number={3},
  pages={225--361},
  year={2013},
  publisher={Taylor \& Francis},
  url={https://www.tandfonline.com/doi/abs/10.1080/00018732.2013.803819}
}

@article{mejia2011first,
title={First passages for a search by a swarm of independent random searchers},
  author={Mej{\'\i}a-Monasterio, Carlos and Oshanin, Gleb and Schehr, Gr{\'e}gory},
  journal={J. Stat. Mech: Theory Exp.},
  volume={2011},
  number={06},
  pages={P06022},
  year={2011},
  publisher={IOP Publishing},
  url={https://iopscience.iop.org/article/10.1088/1742-5468/2011/06/P06022}
}

@article{bhat2016stochastic,
title={Stochastic search with Poisson and deterministic resetting},
  author={Bhat, Uttam and De Bacco, Caterina and Redner, S},
  journal={J. Stat. Mech: Theory Exp.},
  volume={2016},
  number={8},
  pages={083401},
  year={2016},
  publisher={IOP Publishing},
  url={https://iopscience.iop.org/article/10.1088/1742-5468/2016/08/083401}
}

@article{larralde1992territory,
title={Territory covered by N diffusing particles},
  author={Larralde, Hernan and Trunfio, Paul and Havlin, Shlomo and Stanley, H Eugene and Weiss, George H},
  journal={Nature},
  volume={355},
  number={6359},
  pages={423--426},
  year={1992},
  publisher={Nature Publishing Group UK London},
  url={https://www.nature.com/articles/355423a0}
}

@article{acedo2003multiparticle,
title={Multiparticle random walks},
  author={Acedo, Luis and Yuste, Santos B},
  journal={arXiv preprint cond-mat/0310121},
  year={2003},
  url={https://arxiv.org/abs/cond-mat/0310121}
}

@article{majumdar2012number,
  title = {Number of common sites visited by $N$ random walkers},
  author = {Majumdar, Satya N. and Tamm, Mikhail V.},
  journal = {Phys. Rev. E},
  volume = {86},
  issue = {2},
  pages = {021135},
  numpages = {5},
  year = {2012},
  month = {Aug},
  publisher = {American Physical Society},
  doi = {10.1103/PhysRevE.86.021135},
  url = {https://link.aps.org/doi/10.1103/PhysRevE.86.021135}
}

@article{kundu2013exact,
  title = {Exact Distributions of the Number of Distinct and Common Sites Visited by $N$ Independent Random Walkers},
  author = {Kundu, Anupam and Majumdar, Satya N. and Schehr, Gr\'egory},
  journal = {Phys. Rev. Lett.},
  volume = {110},
  issue = {22},
  pages = {220602},
  numpages = {5},
  year = {2013},
  month = {May},
  publisher = {American Physical Society},
  doi = {10.1103/PhysRevLett.110.220602},
  url = {https://link.aps.org/doi/10.1103/PhysRevLett.110.220602}
}

@article{mookerjee2025closed,
title={Closed-form survival probabilities for biased random walks at arbitrary step number},
  author={Mookerjee, Debendro and Kostinski, Sarah},
  journal={Phys. Rev. Res.},
  volume={7},
  number={4},
  pages={L042007},
  year={2025},
  publisher={APS},
  url={https://doi.org/10.1103/krxn-vqnt}
}

@article{chakraborty2007finite,
title={Finite-size effect in persistence in random walks},
  author={Chakraborty, D and Bhattacharjee, JK},
  journal={Phys. Rev. E: Stat. Nonlinear Soft Matter Phys.},
  volume={75},
  number={1},
  pages={011111},
  year={2007},
  publisher={APS},
  url={https://doi.org/10.1103/PhysRevE.75.011111}
}

@article{chakraborty2008persistence,
title={Persistence in random walk in composite media},
  author={Chakraborty, D},
  journal={The European Physical Journal B},
  volume={64},
  number={2},
  pages={263--269},
  year={2008},
  publisher={Springer},
  url={https://link.springer.com/article/10.1140/epjb/e2008-00300-1#citeas}
}

@article{majumdar2017survival,
  title={Survival probability of random walks and L{\'e}vy flights on a semi-infinite line},
  author={Majumdar, Satya N and Mounaix, Philippe and Schehr, Gr{\'e}gory},
  journal={J. Phys. A: Math. Theor.},
  volume={50},
  number={46},
  pages={465002},
  year={2017},
  publisher={IOP Publishing},
   url={https://iopscience.iop.org/article/10.1088/1751-8121/aa8d28/meta}
}

@article{turban2014probability,
title={Probability distribution of the number of distinct sites visited by a random walk on the finite-size fully-connected lattice},
  author={Turban, Lo{\"\i}c},
  journal={J. Phys. A: Math. Theor.},
  volume={47},
  number={38},
  pages={385004},
  year={2014},
  publisher={IOP Publishing},
  url={https://iopscience.iop.org/article/10.1088/1751-8113/47/38/385004}
}

@article{evans2020stochastic,
title={Stochastic resetting and applications},
  author={Evans, Martin R and Majumdar, Satya N and Schehr, Gr{\'e}gory},
  journal={J. Phys. A: Math. Theor.},
  volume={53},
  number={19},
  pages={193001},
  year={2020},
  publisher={IOP Publishing},
  url={https://iopscience.iop.org/article/10.1088/1751-8121/ab7cfe}
}

@article{kusmierz2014first,
title={First Order Transition for the Optimal Search Time of Lévy Flights with Resetting},
  author={Kusmierz, Lukasz and Majumdar, Satya N and Sabhapandit, Sanjib and Schehr, Gr{\'e}gory},
  journal={Phys. Rev. Lett.},
  volume={113},
  number={22},
  pages={220602},
  year={2014},
  publisher={APS},
  url={https://doi.org/10.1103/PhysRevLett.113.220602}
}

@article{pal2015diffusion,
title={Diffusion in a potential landscape with stochastic resetting},
  author={Pal, Arnab},
  journal={Phys. Rev. E},
  volume={91},
  number={1},
  pages={012113},
  year={2015},
  publisher={APS},
  url={https://doi.org/10.1103/PhysRevE.91.012113}
}

@article{campos2015phase,
title={Phase transitions in optimal search times: How random walkers should combine resetting and flight scales},
  author={Campos, Daniel and M{\'e}ndez, Vicen{\c{c}}},
  journal={Phys. Rev. E},
  volume={92},
  number={6},
  pages={062115},
  year={2015},
  publisher={APS},
  url={https://doi.org/10.1103/PhysRevE.92.062115}
}

@article{pal2016diffusion,
title={Diffusion under time-dependent resetting},
  author={Pal, Arnab and Kundu, Anupam and Evans, Martin R},
  journal={J. Phys. A: Math. Theor.},
  volume={49},
  number={22},
  pages={225001},
  year={2016},
  publisher={IOP Publishing},
  url={https://iopscience.iop.org/article/10.1088/1751-8113/49/22/225001}
}

@article{reuveni2016optimal,
title={Optimal Stochastic Restart Renders Fluctuations in First Passage Times Universal},
  author={Reuveni, Shlomi},
  journal={Phys. Rev. Lett.},
  volume={116},
  number={17},
  pages={170601},
  year={2016},
  publisher={APS},
  url={https://doi.org/10.1103/PhysRevLett.116.170601}
}

@article{evans2018run,
title={Run and tumble particle under resetting: a renewal approach},
  author={Evans, Martin R and Majumdar, Satya N},
  journal={J. Phys. A: Math. Theor.},
  volume={51},
  number={47},
  pages={475003},
  year={2018},
  publisher={IOP Publishing},
  url={https://iopscience.iop.org/article/10.1088/1751-8121/aae74e}
}

@article{nagar2023stochastic,
  title={Stochastic resetting in interacting particle systems: A review},
  author={Nagar, Apoorva and Gupta, Shamik},
  journal={J. Phys. A: Math. Theor.},
  volume={56},
  number={28},
  pages={283001},
  year={2023},
  publisher={IOP Publishing},
  url={https://iopscience.iop.org/article/10.1088/1751-8121/acda6c}
}

@article{reuveni2014role,
  title={The role of substrate unbinding in michaelis-menten enzymatic reactions},
  author={Reuveni, Shlomi and Urbakh, Michael and Klafter, Joseph},
  journal={Biophys. J.},
  volume={106},
  number={2},
  pages={677a},
  year={2014},
  publisher={Elsevier},
  url={https://doi.org/10.1016/j.bpj.2013.11.3751}
}

@article{rotbart2015michaelis,
title={Michaelis-Menten reaction scheme as a unified approach towards the optimal restart problem},
  author={Rotbart, Tal and Reuveni, Shlomi and Urbakh, Michael},
  journal={Phys. Rev. E},
  volume={92},
  number={6},
  pages={060101},
  year={2015},
  publisher={APS},
  url={https://doi.org/10.1103/PhysRevE.92.060101}
}

@article{montanari2002optimizing,
title={Optimizing Searches via Rare Events},
  author={Montanari, Andrea and Zecchina, Riccardo},
  journal={Phys. Rev. Lett.},
  volume={88},
  number={17},
  pages={178701},
  year={2002},
  publisher={APS},
  url={https://doi.org/10.1103/PhysRevLett.88.178701}
}

@article{boyer2014random,
title={Random Walks with Preferential Relocations to Places Visited in the Past and their Application to Biology},
  author={Boyer, Denis and Solis-Salas, Citlali},
  journal={Phys. Rev. Lett.},
  volume={112},
  number={24},
  pages={240601},
  year={2014},
  publisher={APS},
  url={https://doi.org/10.1103/PhysRevLett.112.240601}
}

@article{pal2020search,
title={Search with home returns provides advantage under high uncertainty},
  author={Pal, Arnab and Ku{\'s}mierz, {\L}ukasz and Reuveni, Shlomi},
  journal={Phys. Rev. Res.},
  volume={2},
  number={4},
  pages={043174},
  year={2020},
  publisher={APS},
  url={https://doi.org/10.1103/PhysRevResearch.2.043174}
}

@article{vilk2022phase,
title={Phase Transition in a Non-Markovian Animal Exploration Model with Preferential Returns},
  author={Vilk, Ohad and Campos, Daniel and M{\'e}ndez, Vicen{\c{c}} and Lourie, Emmanuel and Nathan, Ran and Assaf, Michael},
  journal={Phys. Rev. Lett.},
  volume={128},
  number={14},
  pages={148301},
  year={2022},
  publisher={APS},
  url={https://doi.org/10.1103/PhysRevLett.128.148301}
}

@article{pal2017first,
title={First Passage under Restart},
  author={Pal, Arnab and Reuveni, Shlomi},
  journal={Phys. Rev. Lett.},
  volume={118},
  number={3},
  pages={030603},
  year={2017},
  publisher={APS},
  url={https://doi.org/10.1103/PhysRevLett.118.030603}
}

@article{tal2020experimental,
title={Experimental Realization of Diffusion with Stochastic Resetting},
  author={Tal-Friedman, Ofir and Pal, Arnab and Sekhon, Amandeep and Reuveni, Shlomi and Roichman, Yael},
  journal={J. Phys. Chem. Lett.},
  volume={11},
  number={17},
  pages={7350--7355},
  year={2020},
  publisher={ACS Publications},
  url={https://doi.org/10.1021/acs.jpclett.0c02122}
}

@article{besga2020optimal,
title={Optimal mean first-passage time for a Brownian searcher subjected to resetting: Experimental and theoretical results},
  author={Besga, Benjamin and Bovon, Alfred and Petrosyan, Artyom and Majumdar, Satya N and Ciliberto, Sergio},
  journal={Phys. Rev. Res.},
  volume={2},
  number={3},
  pages={032029},
  year={2020},
  publisher={APS},
  url={https://doi.org/10.1103/PhysRevResearch.2.032029}
}

@article{faisant2021optimal,
title={Optimal mean first-passage time of a Brownian searcher with resetting in one and two dimensions: experiments, theory and numerical tests},
  author={Faisant, Felix and Besga, Benjamin and Petrosyan, Artyom and Ciliberto, Sergio and Majumdar, Satya N},
  journal={J. Stat. Mech: Theory Exp.},
  volume={2021},
  number={11},
  pages={113203},
  year={2021},
  publisher={IOP Publishing},
  url={https://iopscience.iop.org/article/10.1088/1742-5468/ac2cc7}
}

@article{cayci2020continuous,
title={Continuous-Time Multi-Armed Bandits with Controlled Restarts},
  author={Cayci, Semih and Eryilmaz, Atilla and Srikant, R},
  journal={arXiv preprint arXiv:2007.00081},
  year={2020},
  url={https://doi.org/10.48550/arXiv.2007.00081}
}

@article{lorenz2021restart,
title={Restart Strategies in a Continuous Setting},
  author={Lorenz, Jan-Hendrik},
  journal={Theory Comput. Syst.},
  volume={65},
  number={8},
  pages={1143--1164},
  year={2021},
  publisher={Springer},
  url={https://link.springer.com/article/10.1007/s00224-021-10041-0}
}

@inproceedings{lorenz2018runtime,
  author={Lorenz, Jan-Hendrik},
  booktitle={SOFSEM 2018: Theory and Practice of Computer Science: 44th International Conference on Current Trends in Theory and Practice of Computer Science, Krems, Austria, January 29-February 2, 2018, Proceedings 44},
  pages={493--507},
  year={2018},
  organization={Springer}
}

@article{ghosh2020persistence,
title={Persistence in Brownian motion of an ellipsoidal particle in two dimensions},
  author={Ghosh, Anirban and Chakraborty, Dipanjan},
  journal={J. Chem. Phys.},
  volume={152},
  number={17},
  year={2020},
  publisher={AIP Publishing},
  url={https://doi.org/10.1063/5.0004134}
}

@article{ghosh2022persistence,
title={Persistence of an active asymmetric rigid Brownian particle in two dimensions},
  author={Ghosh, Anirban and Mandal, Sudipta and Chakraborty, Dipanjan},
  journal={J. Chem. Phys.},
  volume={157},
  number={19},
  year={2022},
  publisher={AIP Publishing},
  url={https://doi.org/10.1063/5.0119081}
}

@Inbook{Pal2024,
author="Pal, Arnab
and Stojkoski, Viktor
and Sandev, Trifce",
editor="Grebenkov, Denis
and Metzler, Ralf
and Oshanin, Gleb",
title="Random Resetting in Search Problems",
bookTitle="Target Search Problems",
year="2024",
publisher="Springer Nature Switzerland",
address="Cham",
pages="323--355",
isbn="978-3-031-67802-8",
doi="10.1007/978-3-031-67802-8_14",
url="https://doi.org/10.1007/978-3-031-67802-8_14"
}

@article{PhysRevE.99.032123,
  author = {Pal, Arnab and Prasad, V. V.},
  journal = {Phys. Rev. E},
  volume = {99},
  issue = {3},
  pages = {032123},
  numpages = {11},
  year = {2019},
  month = {Mar},
  publisher = {American Physical Society},
  doi = {10.1103/PhysRevE.99.032123},
  url = {https://link.aps.org/doi/10.1103/PhysRevE.99.032123}
}

@article{PhysRevE.103.052129,
  author = {Bonomo, Ofek Lauber and Pal, Arnab},
  journal = {Phys. Rev. E},
  volume = {103},
  issue = {5},
  pages = {052129},
  numpages = {14},
  year = {2021},
  month = {May},
  publisher = {American Physical Society},
  doi = {10.1103/PhysRevE.103.052129},
  url = {https://link.aps.org/doi/10.1103/PhysRevE.103.052129}
}

@article{ghosh2026anisotropic,
title={Anisotropic active Brownian particle in two dimensions under stochastic resetting},
  author={Ghosh, Anirban and Mandal, Sudipta and Chaki, Subhasish},
  journal={Phys. Rev. E},
  volume={113},
  number={1},
  pages={014142},
  year={2026},
  publisher={APS},
  url={https://doi.org/10.1103/11f6-srsx}
}

@article{paramanick2024uncovering,
title={Uncovering Universal Characteristics of Homing Paths using Foraging Robots},
  author={Paramanick, Somnath and Biswas, Arup and Soni, Harsh and Pal, Arnab and Kumar, Nitin},
  journal={Phys. Rev. X Life},
  volume={2},
  number={3},
  pages={033007},
  year={2024},
  publisher={APS},
  url={https://doi.org/10.1103/PRXLife.2.033007}
}

@article{evans2011diffusion,
title={Diffusion with Stochastic Resetting},
  author={Evans, Martin R and Majumdar, Satya N},
journal={Phys. Rev. Lett.},
  volume={106},
  number={16},
  pages={160601},
  year={2011},
  publisher={APS},
  url={https://doi.org/10.1103/PhysRevLett.106.160601}
}

@article{PhysRevLett.116.170601,
  author = {Reuveni, Shlomi},
  journal = {Phys. Rev. Lett.},
  volume = {116},
  issue = {17},
  pages = {170601},
  numpages = {6},
  year = {2016},
  month = {Apr},
  publisher = {American Physical Society},
  doi = {10.1103/PhysRevLett.116.170601},
  url = {https://link.aps.org/doi/10.1103/PhysRevLett.116.170601}
}

@article{majumdar1999persistence,
title={Persistence in nonequilibrium systems},
  author={Majumdar, Satya N},
  journal={Current Science},
  pages={370--375},
  year={1999},
  publisher={JSTOR},
  url={https://www.jstor.org/stable/24102955}
}

@article{linn2025cover,
title={Cover times with stochastic resetting},
  author={Linn, Samantha and Lawley, Sean D},
  journal={Chaos: An Interdisciplinary Journal of Nonlinear Science},
  volume={35},
  number={4},
  year={2025},
  publisher={AIP Publishing},
  url={https://doi.org/10.1063/5.0260643}
}

@article{kundu2024preface,
title={Preface: stochastic resetting—theory and applications},
  author={Kundu, Anupam and Reuveni, Shlomi},
  journal={J. Phys. A: Math. Theor.},
  number={6},
  pages={060301},
  year={2024},
  publisher={IOP Publishing},
  url={https://iopscience.iop.org/article/10.1088/1751-8121/ad1e1b}
}

@article{bressloff2020queueing,
title={Queueing theory of search processes with stochastic resetting},
  author={Bressloff, Paul C},
  journal={Phys. Rev. E},
  volume={102},
  number={3},
  pages={032109},
  year={2020},
  publisher={APS},
  url={https://doi.org/10.1103/PhysRevE.102.032109}
}

@article{santra2022effect,
title={Universal framework for the long-time position distribution of free active particles},
  author={Santra, Ion and Basu, Urna and Sabhapandit, Sanjib},
  journal={J. Phys. A: Math. Theor.},
  volume={55},
  number={41},
  pages={414002},
  year={2022},
  publisher={IOP Publishing},
  url={https://iopscience.iop.org/article/10.1088/1751-8121/ac864c}
}

@article{meylahn2015large,
title={Large deviations for Markov processes with resetting},
  author={Meylahn, Janusz M and Sabhapandit, Sanjib and Touchette, Hugo},
  journal={Phys. Rev. E},
  volume={92},
  number={6},
  pages={062148},
  year={2015},
  publisher={APS},
  url={https://doi.org/10.1103/PhysRevE.92.062148}
}

@article{majumdar2015dynamical,
title={Dynamical transition in the temporal relaxation of stochastic processes under resetting},
  author={Majumdar, Satya N and Sabhapandit, Sanjib and Schehr, Gr{\'e}gory},
  journal={Phys. Rev. E},
  volume={91},
  number={5},
  pages={052131},
  year={2015},
  publisher={APS},
  url={https://doi.org/10.1103/PhysRevE.91.052131}
}

@article{majumdar2022record,
title={Record statistics for random walks and L{\'e}vy flights with resetting},
  author={Majumdar, Satya N and Mounaix, Philippe and Sabhapandit, Sanjib and Schehr, Gregory},
  journal={J. Phys. A: Math. Theor.},
  volume={55},
  number={3},
  pages={034002},
  year={2022},
  publisher={IOP Publishing},
  url={https://iopscience.iop.org/article/10.1088/1751-8121/ac3fc1/meta}
}

@article{santra2020run,
title={Run-and-tumble particles in two dimensions under stochastic resetting conditions},
  author={Santra, Ion and Basu, Urna and Sabhapandit, Sanjib},
  journal={J. Stat. Mech: Theory Exp.},
  volume={2020},
  number={11},
  pages={113206},
  year={2020},
  publisher={IOP Publishing and SISSA},
  url={https://iopscience.iop.org/article/10.1088/1742-5468/abc7b7}
}

@article{del2026proxitaxis,
title={Proxitaxis: An adaptive search strategy based on proximity and stochastic resetting},
  author={Del Vecchio, Giuseppe Del Vecchio and Kulkarni, Manas and Majumdar, Satya N and Sabhapandit, Sanjib},
  journal={Phys. Rev. E},
  volume={113},
  number={4},
  pages={L042101},
  year={2026},
  publisher={APS},
  url={https://journals.aps.org/pre/abstract/10.1103/bjl2-kmrt}
}

@article{flaquer2024intermittent,
title={Intermittent random walks under stochastic resetting},
  author={Flaquer-Galm{\'e}s, Rosa and Campos, Daniel and M{\'e}ndez, Vicen{\c{c}}},
  journal={Phys. Rev. E},
  volume={109},
  number={3},
  pages={034103},
  year={2024},
  publisher={APS},
  url={https://doi.org/10.1103/PhysRevE.109.034103}
}


\stepcounter{myequation}
\stepcounter{myfigure}

\onecolumngrid

\newpage

\appendix

\section{Calculation of Probability distribution function}\label{app}
From Eqs.~(\ref{saa}) and \eqref{sar}, we are aiming to find out the inverse Laplace transform of $\Tilde{S}^r_{AA}(s,r)$ and $\Tilde{S}^r_{AR}(s,r)$ as

\begin{equation}
    \begin{split}
      &  S^r_{AA}(t)=\frac{1}{2\pi i}\int_{-i\infty}^{i\infty}ds e^{st}\frac{\cosh{\sqrt{\frac{L^2(r+s)}{4D}}}-1}{s\cosh{\sqrt{\frac{L^2(r+s)}{4D}}}+r}\\
      &  S^r_{AR}(t)=\frac{1}{2\pi i}\int_{-i\infty}^{i\infty}ds e^{st}\frac{\cosh{2\sqrt{\frac{L^2(r+s)}{4D}}}-\cosh{\sqrt{\frac{L^2(r+s)}{4D}}}}{s\cosh{2\sqrt{\frac{L^2(r+s)}{4D}}}+r\cosh{\sqrt{\frac{L^2(r+s)}{4D}}}}
    \end{split}
\end{equation}
Let us now consider $\frac{L^2(r+s)}{4D}=w$, $\frac{4Dt}{L^2}=z$, and $\frac{L^2r}{4D}=\rho$, the simplified form of the above expression becomes
\begin{equation}
    \begin{split}
        & S^r_{AA}=\frac{e^{-\rho z}}{2\pi i}\int_{-i\infty}^{i\infty}dw \frac{e^{wz}\Big(\cosh{\sqrt{w}}-1\Big)}{(w-\rho)\cosh{\sqrt{w}}+\rho}\\
        &S^r_{AR}=\frac{e^{-\rho z}}{2\pi i }\int_{-i\infty}^{i\infty}dw \frac{e^{w z}\Big(\cosh{2\sqrt{w}}-\cosh{\sqrt{w}}\Big)}{(w-\rho)\cosh{2\sqrt{w}}+\rho \cosh{\sqrt{w}}}
    \end{split}
\end{equation}

To evaluate the above integrals, we use the residue method. It is evident that $w=0$ is not a pole, for any values of $\rho$. Moreover, all the poles in both integrals are on the negative $w$ side. 
Let us consider that for any generic value of $\rho$, the negatives poles closest to the origin are, respectively, $w_1(\rho)$ and $w_2(\rho)$. Then the leading order expressions for both $S_{AA}^r$ and $S_{AR}^r$ for large $z$, can be found calculating residues at $w_1(\rho)$ and $w_2(\rho)$ respectively. This yields,
\begin{equation}
    \begin{split}      &S_{AA}^r(z)\simeq \frac{2\sqrt{w_1}\Big(\cosh{(\sqrt{w_1})}-1\Big)}{2\sqrt{w_1}\cosh{(\sqrt{w_1})}+(w_1-\rho)\sinh{(\sqrt{w_1}})}e^{-(\rho-w_1)z}=A_1(\rho)e^{-B_1(\rho)z}\\
        &S_{AR}^r(z)\simeq \frac{2\sqrt{w_2}\Big(\cosh{(2\sqrt{w_2})}-\cosh{(\sqrt{w_2})}\Big)}{\rho\sinh{(\sqrt{w_2})}+2(w_2-\rho)\sinh{(2\sqrt{w_2})}+2\sqrt{w_2}\cosh{(2\sqrt{w_2})}}e^{-(\rho-w_2)z}=A_2(\rho)e^{-B_2(\rho)z}
    \end{split}
    \label{A3}
\end{equation}

We find the poles of the above complex Bromwich integrals by numerically solving the denominator for zero. For example, for $\rho=\rho^*$, we find the poles for $S_{AA}^r$ at $w_1*=-0.277064$ and for $S_{AR}^r$ at $w_2^*=1.07391$.
Calculating the residues the final values of $S_{AR}^r(z)$ and $S_{AA}^r(z)$ for $\rho=\rho^*$ become
\begin{equation}
\begin{split}
&S_{AA}^r(z)=1.2068 e^{-2.046854 z}\\
 &   S_{AR}^r(z)=0.994546e^{-0.69585z}
    \end{split}
    \label{saar}
\end{equation}
The derivative of both $S_{AA}^r(z)$ and $S_{AR}^r(z)$ with respect to $z$ is found as
\begin{equation}
\begin{split}
& S_{AA}^{r\prime}(z)=-2.47014e^{-2.046854z}\\
&     S_{AR}^{r\prime}(z)=-0.692055e^{-0.69585z}
\end{split}
\label{scale}
\end{equation}


\section{Calculation of PDF when $z\rightarrow 0$}\label{appB}

---

To compute the survival probabilities for $z \rightarrow 0$, we use the identity
\begin{equation}
    \frac{1}{\cosh{p\sqrt{w}}}=2\sum_{n=0}^{\infty}(-1)^ne^{-(2n+1)p\sqrt{w}}
\end{equation}
in Eqs.~(\ref{saa}) and (\ref{sar}), and get
\begin{equation}
    \begin{split}
        &\Tilde{S}^r_{AA}(s)=\frac{1-2\sum_{n=0}^{\infty}(-1)^ne^{-(2n+1)\sqrt{\frac{L^2(r+s)}{4D}}}}{s+2r\sum_{n=0}^{\infty}(-1)^ne^{-(2n+1)\sqrt{\frac{L^2(r+s)}{4D}}}}\\
        &\Tilde{S}^r_{AR}(s)=\frac{\sum_{n=0}^{\infty}(-1)^ne^{-(2n+1)\sqrt{\frac{L^2(r+s)}{4D}}}-\sum_{n=0}^{\infty}(-1)^ne^{-(2n+1)2\sqrt{\frac{L^2(r+s)}{4D}}}}{s\sum_{n=0}^{\infty}(-1)^ne^{-(2n+1)\sqrt{\frac{L^2(r+s)}{4D}}}+r\sum_{n=0}^{\infty}(-1)^ne^{-(2n+1)2\sqrt{\frac{L^2(r+s)}{4D}}}}
    \end{split}
    \label{39}
\end{equation}

To obtain the survival probabilities \( S_{AA}^r(t) \) and \( S_{AR}^r(t) \), we must compute 
the inverse Laplace transforms of \( \Tilde{S}_{AA}^r(s) \) and \( \Tilde{S}_{AR}^r(s) \). With change of variables where $w=\frac{L^2(r+s)}{4D}$,  $z=\frac{4Dt_c}{L^2}$ and $\rho=\frac{L^2r}{4D}$, it is easy to see that the scaled survival probabilities are given by 
\begin{equation}
    \begin{split}
     &   S_1^r(z)=\frac{1}{2\pi i}e^{-z\rho}\int_{-i\infty}^{+i\infty}dw e^{wz}\frac{1-2\sum_{n=0}^{\infty}(-1)^ne^{-(2n+1)\sqrt{w}}}{w-\rho+2\rho\sum_{n=0}^{\infty}(-1)^ne^{-(2n+1)\sqrt{w}}}\\
     &S_{2}^r(z)=\frac{1}{2\pi i}e^{-z\rho}\int_{-i\infty}^{+i\infty}dw e^{wz}\frac{\sum_{n=0}^{\infty}(-1)^ne^{-(2n+1)\sqrt{w}}-\sum_{n=0}^{\infty}(-1)^ne^{-(2n+1)2\sqrt{w}}}{(w-\rho)\sum_{n=0}^{\infty}(-1)^ne^{-(2n+1)\sqrt{w}}+\rho \sum_{n=0}^{\infty}(-1)^ne^{-(2n+1)2\sqrt{w}}}   
    \end{split}
\end{equation}
 The leading order behavior for small $z$ can be obtained by considering the $n=0$ in the integrals, which gives,
\begin{equation}
    \begin{split}
     &   S_{1}^r(z)\simeq\frac{1}{2\pi i}e^{-z\rho}\int_{-i\infty}^{+i\infty}dw e^{wz}\frac{1-2e^{-\sqrt{w}}}{w-\rho+2\rho e^{-\sqrt{w}}}\\
     &S_{2}^2(z,\rho)\simeq\frac{1}{2\pi i}e^{-z\rho}\int_{-i\infty}^{+i\infty}dw e^{wz}\frac{e^{-\sqrt{w}}-e^{-2\sqrt{w}}}{(w-\rho)e^{-\sqrt{w}}+\rho e^{-2\sqrt{w}}}  
    \end{split}
    \label{B2}
\end{equation}
If we consider the substitution of variable as $u=wz$ then Eq.(\ref{B2}) can be rewritten as by replacing the variable $w$ in terms of new variable $u$:

\begin{equation}
    \begin{split}
         &   S_{AA}^r(z)=\frac{1}{2\pi i}e^{-z\rho}\int_{-i\infty}^{+i\infty}du e^u\frac{1-2e^{-\sqrt{u/z}}}{u-\rho z+2\rho z e^{-\sqrt{u/z}}}\\
         &   S_{AR}^r(z)=\frac{1}{2\pi i}e^{-z\rho}\int_{-i\infty}^{+i\infty}du e^u\frac{1-e^{-\sqrt{u/z}}}{u-\rho z+\rho z e^{-\sqrt{u/z}}}
    \end{split}
\end{equation}
For $z\rightarrow 0$, $u/z$ becomes large, and consequently,  we ignore the term $e^{-\sqrt{u/z}}$ in the denominator, which gives
\begin{equation}
    \begin{split}
         &   S_{1}^r(z)=\frac{1}{2\pi i}e^{-z\rho}\int_{-i\infty}^{+i\infty}du e^u\frac{1-2e^{-\sqrt{u/z}}}{u-\rho z}\\
         &   S_{2}^r(z)=\frac{1}{2\pi i}e^{-z\rho}\int_{-i\infty}^{+i\infty}du e^u\frac{1-e^{-\sqrt{u/z}}}{u-\rho z}
    \end{split}
    \label{B4}
\end{equation}

The first integral in Eq.(\ref{B4}) can be written in two parts as: first the contribution due to the residue of the pole $u=\rho z$ and second is the contribution due to the branch cut when $\rho, z\geq 0$. The solution becomes,

\begin{equation}
    \begin{split}
        &   S_{1}^r(z)=1-2e^{-\sqrt{\rho}}+\frac{4}{\pi}e^{-z\rho}\int_{0}^{\infty}\frac{e^{-zt^2}t\sin{t}}{t^2+\rho}dt\\
         &S_{2}^r(z)=1-e^{-\sqrt{\rho}}+\frac{2}{\pi}e^{-z\rho}\int_{0}^{\infty}\frac{e^{-zt^2}t\sin{t}}{t^2+\rho}dt
    \end{split}
\end{equation}

The integral $\frac{4}{\pi}e^{-z\rho}\int_{0}^{\infty}\frac{e^{-zt^2}t\sin{t}}{t^2+\rho}dt$ does not have any closed form solution for finite reset rate $\rho$. Proceeding with higher values of $n$ we get generalized expression for both survival probabilities for the condition $z\rightarrow 0$ as:

\begin{equation}
    \begin{split}
          S_{1}^r(z)&=1-2\sum_{n=0}^{\infty}(-1)^ne^{-(2n+1)\sqrt{\rho}}+\frac{4}{\pi}\sum_{n=0}^{\infty}(-1)^ne^{-z\rho}\int_{0}^{\infty}\frac{e^{-zt^2}t\sin{(2n+1)t}}{t^2+\rho}dt\\
        &=1-\operatorname{sech}\sqrt{\rho}+\frac{4}{\pi}\sum_{n=0}^{\infty}(-1)^ne^{-z\rho}\int_{0}^{\infty}\frac{e^{-zt^2}t\sin{(2n+1)t}}{t^2+\rho}dt
    \end{split}
    \label{full_soln}
\end{equation}
\begin{equation}
     S_{2}^r(z)=1-\sum_{n=0}^{\infty}\bigg(\sin{\frac{n\pi}{2}}+\cos{\frac{n\pi}{2}}\bigg)e^{-(2n+1)\sqrt{\rho}}+\frac{2}{\pi}\sum_{n=0}^{\infty}\bigg(\sin{\frac{n\pi}{2}}+\cos{\frac{n\pi}{2}}\bigg)e^{-z\rho}\int_{0}^{\infty}\frac{e^{-zt^2}t\sin{(2n+1)t}}{t^2+\rho}dt
     \label{full_soln_S}
\end{equation}

If we consider no reset condition at $\rho\rightarrow 0$ in Eq.(\ref{full_soln}) and (\ref{full_soln_S}), the results of Ref.\cite{majumdar2016exact} are found considering  $erf(\frac{2n+1}{2\sqrt{z}})=\frac{2}{\pi}\int_{0}^{\infty}\frac{e^{-zt^2}\sin{(2n+1)t}}{t}dt$. Consequently, the probability distribution function (PDF) scaling function can be written as $f_1^r(z,\rho)=\frac{\partial S_{AA}^{r}(z,\rho)}{\partial z}-2\frac{\partial S_{AR}^{r}(z,\rho)}{\partial z}$. Only the third term of expressions Eq.(\ref{full_soln}) and (\ref{full_soln_S}) will contribute when they are differentiated with respect to $z$, other parts will readily contribute to zero. The third terms in both these expressions have a common integral $e^{-z\rho}\int_{0}^{\infty}\frac{e^{-zt^2}t\sin{(2n+1)t}}{t^2+\rho}dt$, which needs to be differentiated with respect to $z$ as follows:

$$
\frac{\partial}{\partial z}\left(e^{-z\rho}\int_{0}^{\infty}\frac{e^{-zt^2}t\sin{(2n+1)t}}{t^2+\rho}dt\right) = -\rho e^{-z\rho}\int_{0}^{\infty}\frac{e^{-zt^2}t\sin{(2n+1)t}}{t^2+\rho}dt+e^{-z\rho}\left(-\int_{0}^{\infty}\frac{t^3e^{-zt^2}\sin{(2n+1)t}}{t^2+\rho}dt\right)
$$
$$
= -e^{-z\rho}\int_{0}^{\infty}\frac{(\rho t+t^3)}{t^2+\rho}e^{-zt^2}\sin{(2n+1)t}dt
$$
$$
= -e^{-z\rho}\int_{0}^{\infty}t e^{-zt^2}\sin{(2n+1)t}dt
$$
The above integral has a known closed-form solution 
$$
\int_{0}^{\infty}t e^{-zt^2}\sin{(2n+1)t}dt=\frac{(2n+1)\sqrt{\pi}}{4z^{3/2}}e^{-\frac{(2n+1)^2}{4z}}
$$

The scaling function for PDF is expressed by using the above results as:
\begin{equation}
    f_1^r(z)=\frac{e^{-z\rho}}{\sqrt{\pi}z^{3/2}}\sum_{n=0}^{\infty}\bigg(\sin{\frac{n\pi}{2}}+\cos{\frac{n\pi}{2}}\bigg)(2n+1)e^{-\frac{(2n+1)^2}{4z}}-\frac{e^{-z\rho}}{\sqrt{\pi}z^{3/2}}\sum_{n=0}^{\infty}(-1)^n(2n+1)e^{-\frac{(2n+1)^2}{4z}}
    \label{distr}
\end{equation}

\end{document}